\documentclass{aa}
\usepackage{graphics}

\begin{document}


\title{\object{65~Cybele} in the thermal infrared: Multiple observations and thermophysical
       analysis\thanks{Partly based on observations with ISO, an ESA project
          with instruments funded by ESA Member States (especially
          the PI countries: France, Germany, the Netherlands
          and the United Kingdom) and with the participation
          of ISAS and NASA.}
       }

\author{T. G. M\"uller\inst{1} \and J. A. D. L. Blommaert\inst{2}}
\authorrunning{M\"uller \& Blommaert}
\offprints{T.\ G.\ M\"uller}

\institute{Max-Planck-Institut f\"ur extraterrestrische Physik,
           Giessenbachstra{\ss}e, 85748 Garching, Germany;
           \email{tmueller@mpe.mpg.de}
           \and
           Katholieke Universiteit Leuven, Instituut voor Sterrenkunde
           Celestijnenlaan 200B, B-3001 Leuven, Belgium;
           \email{joris.blommaert@ster.kuleuven.ac.be}}
     
\date{Received / Accepted; compilation date: \today}

\abstract{
  We investigated the physical and thermal properties of \object{65~Cybele},
  one of the largest main-belt asteroids. Based on published and
  recently obtained thermal infrared observations, including ISO
  measurements, we derived through thermophysical modelling (TPM)
  a size of 302$\times$290$\times$232\,km ($\pm$ 4\,\%) and
  an geometric visible albedo of 0.050$\pm$0.005. Our model of a regolith covered
  surface with low thermal inertia and ``default" roughness
  describes the wavelengths and phase angle dependent thermal
  aspects very well. Before/after opposition effect and beaming
  behaviour can be explained in that way. We found a constant
  emissivity of 0.9 at wavelengths up to about 100\,$\mu$m and
  lower values towards the submillimetre range, indicating a
  grain size distribution dominated by 200\,$\mu$m particle sizes.
  The spectroscopic analysis revealed an emissivity increase between
  8.0 and 9.5\,$\mu$m. We compared this emissivity behaviour 
  with the Christiansen features of carbonaceous chondrite meteorites,
  but a conclusive identification was not possible.
  A comparison between the Standard Thermal Model (STM) and the 
  applied TPM clearly demonstrates the 
  limitations and problems of the STM for the analysis of
  multi-epoch and -wavelengths observations. While the TPM produced
  a unique diameter/albedo solution, the calculated STM values
  varied by $\pm$30\,\% and showed clear trends with wavelength
  and phase angle. \object{65~Cybele} can be
  considered as a nice textbook case for the thermophysical
  analysis of combined optical and thermal infrared observations.
}
\maketitle

\section{Introduction} \label{sec:intro}

  The asteroid \object{65~Cybele} is a main-belt asteroid
  and the main representative of a group of
  asteroids with semi-major axis in the range 3.27$<$a$<$3.70\,AU,
  eccentricities e$\le$0.30 and inclinations of i$\le$25$^{\circ}$
  (Gradie et al.\ \cite{gradie89}).
  \object{65~Cybele} is generally considered as one of the 10 largest asteroids,
  with a published diameter between 230 (Dobrovol'skij \cite{dobrovol80};
  Taylor \cite{taylor81}) and about 330\,km (Green et al.\ \cite{green85}).
  Although discovered in 1863, it took until 1980 before its rotational
  properties were determined as one of the last remaining large minor
  planets (Schober et al.\ \cite{schober80}). Recent analysis of
  many lightcurve observations lead to a rotation period of about 6\,hours,
  a unique spin vector solution pair with a retrograde sense of rotation
  and an elongated body shape (Erikson \cite{erikson00}).
  
  Based on its low albedo and the flat
  to slightly reddish spectrum in the 0.3 to 1.1\,$\mu$m range, it
  was classified as P-type object (Tholen \cite{tholen89}) and
  due to a weak ultraviolet absorption feature short ward of 0.4\,$\mu$m
  as C0-type (Barucci et al.\ \cite{barucci87}).
  Recently Bus (\cite{bus99}) classified \object{65~Cybele} as Xc type,
  based on reflectance spectra covering the wavelength interval from 0.44
  to 0.92\,$\mu$m. These taxonomic types indicate hydrated silicates, carbon
  and organics on the asteroids surface (Gaffey et al.\ \cite{gaffey89}).
  
  Another interesting aspect comes from the occultation measurement:
  Dobrovol'skij (\cite{dobrovol80}) reported a second event during
  the occultation of AGK3~+19\,599 which
  was attributed to a satellite of 11\,km diameter, located 917\,km
  from the center of \object{65~Cybele} (in the case of a central occultation).
  So far, no confirmation of the satellite through
  other techniques was published. It was one of the targets in the HST
  search for binaries `Imaging Snapshots of Asteroids'
  Storrs et al.\ (\cite{storrs99}) \& Storrs (priv.\ comm.), but with
  unknown outcome so far. No radar measurements are available yet.
 
  Tedesco \& Veeder (\cite{tedesco92}) and Tedesco et al.\
  (\cite{tedesco02}) derived from IRAS observations
  diameter and albedo values of
  237.26$\pm$4.2\,km and 0.0706$\pm$0.003, respectively.
  The only direct size information from the occultation event
  in 1979 resulted in a diameter of 230$\pm$16\,km, with a
  largest measured chord of 245\,km (Dobrovol'skij \cite{dobrovol80};
  Taylor \cite{taylor81}; Xiu-Yi \cite{xiuyi79}).
  Both values, the radiometric
  and the occultation diameter, agree nicely.
  But ground-based N-band observations (Morrison \& Chapman \cite{morrison76};
  Morrison \cite{morrison77}; Bowell et al. \cite{bowell79};
  Green et al.\ \cite{green85}) lead to diameters larger than 300\,km. 
  The published radiometric diameters and albedos were all
  based on the concept of the Standard Thermal Model (STM;
  Lebofsky et al.~\cite{lebofsky86}; Lebofsky \& Spencer \cite{lebofsky89}),
  but using different values for the infrared beaming
  parameter $\eta$, the phase angle correction and the emissivity.
  Using for example the STM $\eta$-value of 0.756 
  (Lebofsky et al.~\cite{lebofsky86}) would
  lower some of the earlier 300\,km diameter values to about 250\,km,
  but the Green et al.\ (\cite{green85}) values would still be close
  to 300\,km. The higher phase angle correction of Bowell et
  al.\ (\cite{bowell79}) would again increase the diamter of all
  measurements taken at large phase angles.

  The diameter is the most fundamental parameter, but in case of
  \object{65~Cybele} it has very large uncertainties, much larger
  than for other asteroids of comparable size. 
  Where is this uncertainty coming from? Are the
  data sets too different or some of them unreliable? What
  role play the STM concept, the simplification in the 
  shape, the missing thermal and rotation effects
  or the various other model input parameters?
  
  The first thermal spectrum of \object{65~Cybele} was published
  by Green et al.\ (\cite{green85}). They did not see any pronounced
  emission features in the 8 to 13\,$\mu$m wavelength range.
  In the years 1996--1998 \object{65~Cybele} was extensively observed by ISO
  with the scientific goal of providing new insights into the chemical
  composition and the mineralogy for a sample of main-belt asteroids.
  Are there any spectral features visible? If yes, what are the meteoritic
  and mineralogic counterparts?

   
  With the availability of a recently established thermophysical
  model (TPM) by Lagerros (\cite{lagerros96}; \cite{lagerros97};
  \cite{lagerros98}) we are now in a position to combine the
  information on the asteroid, its rotational behaviour, different
  observing and illumination geometries with the thermal measurements
  taken at different wavelengths and different times.
  
  In the following, we show the results of the re-analysis and re-calibration
  of the
  existing thermal observations from ground- and satellite-based instruments, 
  including photometric and spectroscopic measurements (Sect.~\ref{sec:obs}).
  In Sect.~\ref{sec:tpm} we present the basic parameters of our
  TPM analysis and derive radiometric diameters and albedos for all
  photometrically reliable data sets (Sect.~\ref{sec:radiometric}).
  In the spectroscopic analysis (Sect.~\ref{sec:spectroscopic}) we focused
  on the modelling of the
  shape of the thermal continuum emission which is strongly connected
  to thermal effects on a regolith covered surface.
  The TPM analysis section is followed by extensive discussions of
  radiometric and spectroscopic results (Sect.~\ref{sec:discussion}).
  We used a large sample of infrared observations, spanning a wide range
  of phase angles and wavelengths, to demonstrate
  the possibilities and limitations of the thermophysical modelling of
  asteroids (Sect.~\ref{sec:conclusion}).
  
\section{Observations in the Thermal Infrared}  \label{sec:obs}
\subsection{Re-analysis of previous measurements}

\paragraph{TRIAD Observations\\}
  The original TRIAD\footnote{Tucson revised Index of Asteroid Data}
  fluxes were based on three \object{65~Cybele}
  observations from March 18, 1975, 08:24 UT, with an average N-band magnitude
  of $+$0.85$\pm$0.05\,mag (Morrison \& Chapman \cite{morrison76}).
  We used the Rieke et al.\ (\cite{rieke85}) absolute photometric system
  for the standard stars to re-calibrate this flux. This led to a new N-band
  value of $+0.92$\,mag, corresponding to a colour corrected flux at
  10.6\,$\mu$m of 14.55$\pm$0.87\,Jy (measurement and calibration errors
  have been added quadratically). 
  
\paragraph{IR Spectroscopy from UCL\\}
  Green et al.\ (\cite{green85}) presented the first 8- to 13\,$\mu$m
  spectrum of \object{65~Cybele}.
  It was taken on April 18, 1982 and calibrated 
  against $\alpha$~Boo. We re-calibrated the
  10.0\,$\mu$m flux value to 25.5$\pm$2.9\,Jy
  with the Cohen et al.\ (\cite{cohen95}) model for $\alpha$~Boo
  (the Green et al.\ value for $\alpha$~Boo is 15.0\,\%
  higher than the Cohen et al.\ value at 10.0\,$\mu$m).
  The re-calibration resulted in flux values of 
  14.4$\pm$2.0\,Jy at 8.5\,$\mu$m, 25.5$\pm$2.9\,Jy at 10.0\,$\mu$m
  and 50.9$\pm$4.4\,Jy at 12.0\,$\mu$m for the given epoch.
  
\paragraph{IRAS Observations\\}

  IRAS (Beichman et al.~\cite{beichman88}) scanned 6 times over
  \object{65~Cybele}, each time with all 4 wavelength bands.
  The latest flux values and the corresponding analysis were presented
  by Tedesco et al.\ (\cite{tedesco02}).
  We used the colour corrected monochromatic IRAS fluxes (without
  applying the STM-related band-to-band corrections) for our reanalysis
  (see also M\"uller \& Lagerros \cite{mueller98}).

\paragraph{IRTF Observations\\}
  We included in our analysis 3 IRTF measurements
  (D.\ Osip, priv.\ communications, see Table~\ref{tbl:irtf_osip}):
  
  \begin{table}[h!tb]
  \begin{center}
    \caption{IRTF observations of \object{65~Cybele} from Dec.\ 19, 1996}
    \label{tbl:irtf_osip}
  \begin{tabular}{ccc}
  \hline
  \noalign{\smallskip}
  Julian Date & $\lambda_c$ [$\mu$m] & Flux [Jy] \\
   \noalign{\smallskip}
   \hline
   \noalign{\smallskip}  
   2450436.71813 & 10.5 &  7.0 \\
   2450436.72042 & 10.5 &  6.7 \\
   2450436.72350 & 18.0	& 17.8 \\
   \noalign{\smallskip}
   \hline
  \end{tabular}
  \end{center}
  \end{table}
  
  No observational or calibration errors were given, but
  the data were considered as a high accuracy dataset (Osip,
  priv.\ comm.). We assumed a 10\,\% flux uncertainty
  for the N-band and 15\,\% for the Q-band photometry.
  
\paragraph{UKIRT Observations\\}
  The UKIRT mid-IR N- and Q-band data of \object{65~Cybele} have been obtained in
  August 1996 with the MPIA 5--25\,$\mu$m camera MAX
  (Robberto \& Herbst \cite{robberto98}). Observations, data 
  reduction and calibration are described in 
  M\"uller \& Lagerros (\cite{mueller98}). 
  
\paragraph{ISOPHOT photometric data\\}
  \object{65~Cybele} was measured at wavelengths between 65 and
  200\,$\mu$m with ISOPHOT (Lemke et al. \cite{lemke96})
  at 3 different observing epochs.
  Observations, data reduction and calibration are discussed in 
  M\"uller \& Lagerros (\cite{mueller98}; \cite{mueller02}).
  
\subsection{ISOPHT PHT-S Measurement}

  \begin{table*}
   \caption[]{PHT-S, CAM CVF and CAM filter observations of \object{65~Cybele}.
              OTT is the On Target Time.}
   \label{tbl:tab1}
   \begin{tabular}{lcccrl}
   \hline
   \noalign{\smallskip}
                 &     & Start       & $\lambda$-range & OTT & \\
      Instrument & TDT & Observation & [$\mu$m] & [sec] & Field of View \\
   \noalign{\smallskip}
   \hline
   \noalign{\smallskip}
      PHT-S                  & 37300324 & 23-Nov-96 12:40:26 &
          5.84--11.62 &  364 & $24^{\prime \prime} \times 24^{\prime \prime}$ \\
   \noalign{\smallskip}
      CAM-CVF forward (down) & 77002401 & 25-Dec-97 02:08:41 &
          17.00--4.96 & 2996 & $96^{\prime \prime} \times 96^{\prime \prime}$ \\
      CAM-Filter             & 77002402 & 25-Dec-97 02:59:19 &
          6.0--14.9  & 3118  & $48^{\prime \prime} \times 48^{\prime \prime}$ \\
      CAM-CVF backward (up)  & 77002403 & 25-Dec-97 03:52:01 &
          5.02--17.00 & 2970 & $96^{\prime \prime} \times 96^{\prime \prime}$ \\
   \noalign{\smallskip}
   \hline
   \end{tabular}
  \end{table*}

  \begin{figure}[btp!]
  \rotatebox{90}{\resizebox{!}{\hsize}{\includegraphics{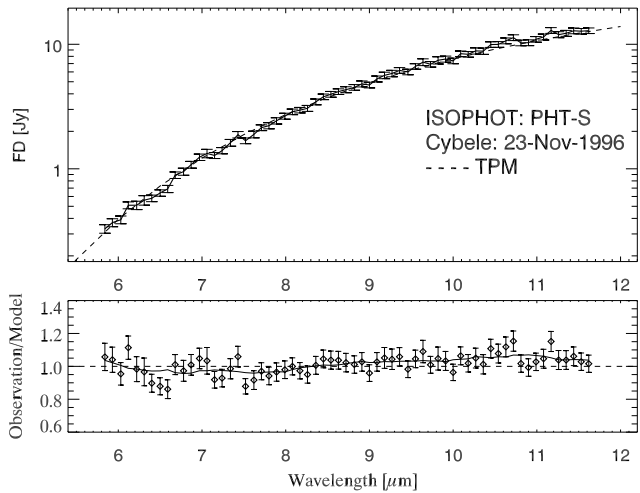}}}
  \caption{PHT-S spectrum of \object{65~Cybele}. The TPM prediction is plotted as dashed
	line (with $D_{eff}=296.5$\,km, $p_V=0.043$ and default thermal parameters
	as described in Sect.~\ref{sec:tpm_input}). The error bars
	are the signal processing errors (between 2 and 6\,\%) added
        quadratically to the assumed calibration error of 5\,\% per pixel. The solid
	line in the lower part was produced through a 10 element smoothing
	of the PHT-S to TPM ratio (see Sect.~\ref{sec:tpm}).
     \label{fig:cyb_phts_tpm}}
  \end{figure}
  
  The observational details of the PHT-S measurements are given
  in Table~\ref{tbl:tab1}. The PHT-S short wavelength part
  (2.47--4.87\,$\mu$m) covers the transition between reflected light
  and thermal emission. It has been excluded in our thermophysical
  analysis. The data reduction of the 5.84--11.62\,$\mu$m part was done in a 
  standard way, using PIA\footnote{ISOPHOT Interactive Analysis}
  (for details of the PHT-S data reduction steps see
  Dotto et al.\ \cite{dotto02}). There was no dedicated background
  measurement, we therefore used COBE-DIRBE measurement results
  (Hauser et al. \cite{hauser98}) as replacement. The color corrected,
  weekly maps from the COBE-DIRBE data sets were adjusted for the
  PHT-S beam size. The two DIRBE values at 4.9 and 12\,$\mu$m were
  then connected using a 266\,K blackbody ({\'A}brah{\'a}m et al.
  \cite{abraham99}) and subtracted from the measured spectrum of
  \object{65~Cybele}. The background contribution was between
  4 and 10\,\% in the PHT-S wavelength range.

  Figure~\ref{fig:cyb_phts_tpm} shows the reduced and background subtracted
  result of the PHT-S observation together with the corresponding
  TPM prediction. Typical signal processing uncertainties were
  between 2 and 6\,\%. The absolute calibration error for bright
  point-like sources in staring mode is less than 10\,\%
  (Laureijs et al. \cite{laureijs02}). For our analysis we added
  quadratically the individual signal processing errors and an absolute
  calibration error of 5\,\%. The lower part of Fig.~\ref{fig:cyb_phts_tpm}
  shows the ratio between the observed spectrum and the TPM prediction
  (see Sect. \ref{sec:tpm}).

\subsection{ISOCAM CVF \& Filter Observations}
\label{sec:cvf}

  \begin{figure}[btp!]
  \rotatebox{90}{\resizebox{!}{\hsize}{\includegraphics{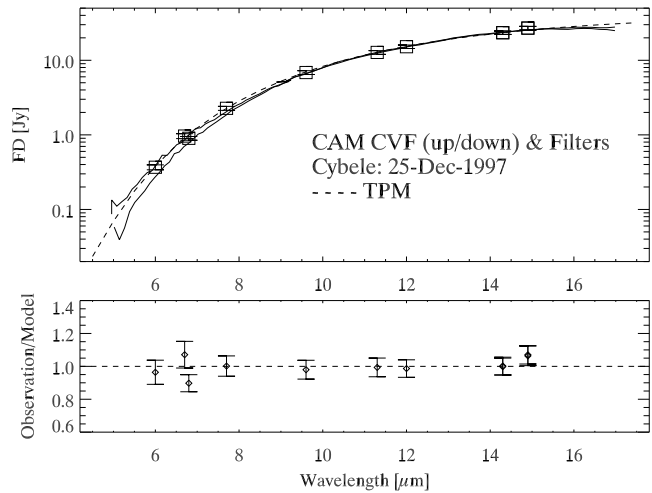}}}
  \caption{CAM-CVF and CAM filter observations of \object{65~Cybele}. The TPM prediction
        is plotted as dashed line (with $D_{eff}=278.8$\,km, $p_V=0.049$ and default
	thermal parameters as described in Sect.~\ref{sec:tpm_input}). The lower
	part shows the resulting CAM filter flux densities divided by the
	corresponding TPM prediction (see Sect.~\ref{sec:tpm}).
     \label{fig:cyb_cam_tpm}}
  \end{figure}
     
  On Christmas day 1997, a for- and backward scan was performed
  in the 5 to 17\,$\mu$m wavelength range with the ISOCAM
  Circular Variable Filter (CVF) (Cesarsky et
  al. \cite{cesarsky96}). The scan was performed skipping one
  CVF position per CVF step, corresponding to steps of approximately
  0.2\,$\mu$m for the 9-17\,$\mu$m and 0.1\,$\mu$m for the
  shorter wavelength range. The pixel field of view was
  3$^{\prime \prime}$ leading to a total field of view of
  96$^{\prime \prime}$. The data reduction was done within the
  CAM Interactive Analysis package (Ott et al. \cite{ott97})
  and followed the same strategy as was done for the standard
  stars observations, used for the calibration of the
  Spectral Response Function (Blommaert et al. \cite{blommaert01}).
  The flux calibration was done using the OLP~v10 SRF\footnote{Off Line
  Processing version 10, Spectral Response Function}.
  Aperture photometry was applied avoiding the contribution
  from reflected light on the detector (Blommaert et al.
  \cite{blommaert01}). The strategy of combining a for- and
  backward scan allows to check observed spectral features and
  also allows to a certain extent the correction of
  the ISOCAM transient response. We applied the Fouks-Schubert
  method (Coulais \& Abergel \cite{coulais00}) to correct the
  transient of the signal.
  The CVF scan starting at 5\,$\mu$m after the fixed filter
  measurements suffers strongly from a remnant which is not
  corrected for by the transient correction method. For this reason
  we judge the data for 5 - 7\,$\mu$m in the up-scan as
  less reliable. 

  In the time between the two CVF scans nine fixed filter
  measurements were performed in the 6--15\,$\mu$m wavelength range.
  The zodiacal background becomes important at wavelengths
  longer than 9\,$\mu$m. Therefore, dedicated background measurements
  were done in the long-wavelength filters. They were performed
  before \object{65~Cybele} itself was observed to avoid remnants of the
  \object{65~Cybele} measurement because of the transients. The on-source
  measurements were long enough to allow stabilisation of the
  signal so that no correction had to be applied (Blommaert
  et al. \cite{blommaert00}). The obtained flux densities
  have been colour corrected using the average CVF as input spectrum.
  
  \begin{table}[h!tb]
  \begin{center}
  \caption{ISOCAM filter measurements. The monochromatic flux values are
           background subtracted and colour corrected. The 1-$\sigma$
	   errors of the photometry are between 1 and 4\,\%.
	   The colour correction terms (Col.~CC) have been
	   determined from the averaged CVF-spectrum.
	   CVF uncertainties produce errors in the colour correction
	   values of about 5\,\% in the LW4 and LW2 filters and
	   about 1\,\% for the others. The uncertainty in
	   flux conversion is approximately 5\,\%.
	   All errors have been added quadratically (Col.~`Error').
	   The measurements were taken on 25 December 1997 between 3:00
	   and 3:50 UT.}
            \label{tbl:iso_camfilter}
  \begin{tabular}{lrcrl}
   \hline
   \noalign{\smallskip}
    filter & $\lambda_{ref}$ [$\mu$m] & CC & Flux [Jy] & Error [Jy] \\
   \noalign{\smallskip}
   \hline
   \noalign{\smallskip}
   LW4  &  6.0 & 1.08 &  0.368 & 0.028 \\
   LW2  &  6.7 & 1.32 &  0.964 & 0.073 \\
   LW5  &  6.8 & 1.00 &  0.899 & 0.052 \\
   LW6  &  7.7 & 1.06 &  2.27  & 0.14 \\
   LW7  &  9.6 & 1.03 &  6.85  & 0.40 \\
   LW8  & 11.3 & 1.01 & 12.74  & 0.73 \\
   LW10 & 12.0 & 0.91 & 15.20  & 0.82 \\
   LW3  & 14.3 & 0.99 & 23.49  & 1.25 \\
        & 14.3 & 0.99 & 23.39  & 1.22 \\
   LW9	& 14.9 & 1.00 & 27.09  & 1.42 \\
   	& 14.9 & 1.00 & 26.92  & 1.50 \\
   \noalign{\smallskip}
   \hline
  \end{tabular}
  \end{center}
  \end{table}
  
  The resulting monochromatic flux densities are listed in
  Table~\ref{tbl:iso_camfilter}. These values are also 
  plotted together with error bars in Fig.~\ref{fig:cyb_cam_tpm}.
  The lower part of the figure
  shows the ratio between the observed CAM filter values and
  the corresponding TPM prediction (see Sect. \ref{sec:tpm}).
  
\subsection{ESO-TIMMI2 N- and Q-band Observations}

  We obtained additional N- and Q-band observations of \object{65~Cybele}
  using the TIMMI2 camera (Reimann et al.\ \cite{reimann00}),
  mounted on the ESO-La Silla 3.6\,m telescope.
  Table~\ref{tbl:timmi2} summarises
  the observational results\footnote{Based on
   observations collected at the European Southern Observatory, Chile;
   ESO, No.\ 69.C-0090}.
  
  \begin{table}[h!tb]
  \begin{center}
  \caption{Summary of the ESO-TIMMI2 N- and Q-band observations.
           All measurements were taken between airmass 1.02 and 1.24.
	   The calibration stars were HD\,178345 and HD\,169916, with
	   model spectra from Cohen et al.\ \cite{cohen99}.}
            \label{tbl:timmi2}
  \begin{tabular}{lrcrrr}
   \hline
   \noalign{\smallskip}
    date & time & filter & $\lambda_{ref}$ & Flux & Error \\
    \multicolumn{2}{c}{(UT)} &  & [$\mu$m] & [Jy] & [Jy] \\
  \noalign{\smallskip}
   \hline
   \noalign{\smallskip}
   10-Jun-02 & 07:42 & Q1 & 17.75 & 74.9  & $\pm^{3}_{15}$ \\
   11-Jun-02 & 08:47 & Q1 & 17.75 & 57.4  &  5.0 \\
             & 08:51 & Q1 & 17.75 & 57.7  &  5.0 \\
	     & 10:02 & N1 &  8.60 &  8.8  &  0.5 \\
   \noalign{\smallskip}
   \hline
  \end{tabular}
  \end{center}
  \end{table}
  
  We did a standard data reduction for the chopping-nodding imaging
  mode. Asteroid and calibration star measurements have been taken
  within 30\,min, all N1 integration time were 144\,sec,
  the Q1 integration times were 108\,sec. The relatively large
  uncertainties in table~\ref{tbl:timmi2} are due to the variable
  atmospheric conditions. The flux errors are estimates based on
  the monitoring of the reference stars. No colour or airmass
  correction was performed, both effects are only on a 1-2\,\%
  level and much smaller than the photometric errors. 
  
\section{Thermophysical Model Analysis}  \label{sec:tpm}

  We applied the thermophysical model (TPM) developed by
  Lagerros (\cite{lagerros96}; \cite{lagerros97}; \cite{lagerros98})
  and extensively used and refined by M\"uller \& Lagerros
  (\cite{mueller98}; \cite{mueller02}). The TPM includes
  shape, rotational behaviour and thermal behaviour of a given
  asteroid at the specific observing and illumination geometry.
  It allows to derive radiometric sizes and albedos from
  thermal infrared observations, taking shape information, geometry
  and thermal aspects into account. 
    
\subsection{Model input values}  \label{sec:tpm_input}
  
  The $H$\footnote{Absolute magnitude in the H-G magnitude system}
  and $G$\footnote{V-band slope parameter in the H-G magnitude system}
  values are taken from Lagerkvist \& Magnusson
  (\cite{lagerkvist90}), with $H=6.70\pm0.15$ as the weighted mean value
  of four apparitions and $G=0.09\pm0.05$.
  Erikson (\cite{erikson00}) presented a good quality solution pair
  for the spin vector based on a retrograde sense of rotation and a
  synodic rotation period of 6.1\,hours (see Table~\ref{tbl:sv}).
  However, due to noisy and under-sampled input data, the spin vector
  pair is still considered only as class~II, indicating that higher
  quality data might improve the situation.

  The sense of rotation can in principle be determined from 
  the anlysis of thermal observations before and after opposition
  (e.g.\ M\"uller \cite{mueller02a}). In cases where two spin vector
  solutions with different senses of rotation are given, the thermophysical
  analysis allows to determine the correct one. But to distinguish between
  two spin vector solutions with the same sense of rotation is very
  difficult on basis of infrared data. The before-/after opposition effect
  is then not very pronounced. Our TPM analysis for \object{65~Cybele} gave
  almost identical diameter and albedo values for both spin vector solutions.
  For the first solution (first line in Table~\ref{tbl:sv}) we obtained
  a slightly smaller standard deviation around the mean diameter and
  albedo values. Therefore we accepted solution 1 for all following
  calculations. The corresponding ellipsoidal shape model solution is
  closely connected to this spin vector solution.

  \begin{table*}[h!tb]
    \caption{Spin vector parameters of \object{65~Cybele}
  	     (Erikson \cite{erikson00}).}
    \label{tbl:sv}
    \begin{tabular}{rlllrrr}
    \hline
    \noalign{\smallskip}
      \multicolumn{1}{c}{Asteroid} &
      \multicolumn{2}{c}{Spin Vector} &
      \multicolumn{1}{c}{Sidereal Period} &
      \multicolumn{1}{c}{Absolute rotational phase} &
      \multicolumn{2}{c}{Ellipsoidal model} \\
      &
      \multicolumn{1}{c}{$\lambda_0$} &
      \multicolumn{1}{c}{$\beta_0$} &
      \multicolumn{1}{c}{$P_{sid}$} &
      \multicolumn{1}{c}{$\gamma_0$} &
      \multicolumn{1}{c}{$a/b$} &
      \multicolumn{1}{c}{$b/c$} \\
    \noalign{\smallskip}
    \hline
    \noalign{\smallskip}
      \object{65~Cybele} &  56$\pm$15 & -25$\pm$15 & 0.2572901$\pm$0.0000004
  	& -38$^{\circ}$ at $T_0=2446420.5$ & 1.04$\pm$0.05 & 1.25$\pm$0.10 \\
  	& 237$\pm$15 & -16$\pm$15 & 0.2572901$\pm$0.0000004
  	& -36$^{\circ}$ at $T_0=2446420.5$ & 1.04$\pm$0.05 & 1.24$\pm$0.10 \\
    \noalign{\smallskip}
    \hline
    \end{tabular}
  \end{table*}

  The determination
  of thermophysical quantities such as thermal inertia, beaming
  parameters or emissivity require large samples of infrared measurements
  (M\"uller \& Lagerros \cite{mueller98}), but for regolith covered,
  large main-belt asteroids, default thermal values seem to work fine
  (M\"uller et al.\ \cite{mueller99}).
  They found a thermal inertia value
  of $\Gamma = 15\,\mathrm{J\,m^{-2}\,s^{-0.5}\,K^{-1}}$ for \object{1~Ceres}
  which they also used successfully for other main-belt asteroids
  (M\"uller \& Lagerros \cite{mueller02}). We adopted this $\Gamma$-value,
  as well as the derived beaming model parameters $f = 0.6$ 
  (fraction of surface covered by craters) and $\rho = 0.7$ (r.m.s.\ of
  the surface slopes). These beaming values correspond roughly to
  the ``default'' STM beaming parameter (Lebofsky et al.~\cite{lebofsky86})
  if the heat conduction is neglected (Lagerros~\cite{lagerros98}).
  For the emissivity we used a constant value
  of $\epsilon = 0.9$ at all wavelengths between the near-IR and 100\,$\mu$m.
  At longer wavelengths we used a wavelength dependent emissivity with
  decreasing values from 0.9 at 100\,$\mu$m to 0.8 at 400\,$\mu$m.

\subsection{Radiometric analysis} \label{sec:radiometric}

  \begin{table*}[h!tb]
  \begin{center}
  \caption{$D_{eff}$ and $p_V$ values, together with the standard deviation
           of all calculated values per instrument (for TRIAD only the flux
	   uncertainty has been taken into account). 
	   The values have been derived through the TPM with default input parameters
	   for shape and thermal behaviour (see~\ref{sec:tpm_input}).}
           \label{tbl:tpm_res}
  \begin{tabular}{lcccc}
   \hline
   \noalign{\smallskip}
   Instr.         & $\lambda$ & Asp.       & $D_{eff}$ & $p_V$ \\
   (\# of Obs)  & [$\mu$m]  & [$^{\circ}$] & [km]      &       \\
   \noalign{\smallskip}
   \hline
   \noalign{\smallskip}
   \multicolumn{5}{l}{Photometry}\\
   \noalign{\smallskip}
   \hline
   \noalign{\smallskip}
   TRIAD (1)    & N-band	&  70	      	&  256.4$\pm$ 7.5  & 0.058$\pm$0.003 \\
   IRAS (24)    & 12,25,60,100	&  68--71     	&  262.4$\pm$14.0  & 0.056$\pm$0.006 \\
   UKIRT (4)    & N- \& Q-band  &  87	        & (329.1$\pm$19.9) & (0.035$\pm$0.004) \\
   IRTF (3)     & N- \& Q-band	&  94	      	&  281.5$\pm$15.7  & 0.048$\pm$0.006 \\
   PHT-P/C (13) & 60...200	&  85--94     	&  269.2$\pm$10.0  & 0.053$\pm$0.004 \\
   CAM Fil (11) & 6.0...14.9	& 144	      	&  278.8$\pm$ 7.0  & 0.049$\pm$0.003 \\
   TIMMI2 (4)   & 8.6, 17.75    &  66           &  260.5$\pm$19.2  & 0.057$\pm$0.008 \\
   \noalign{\smallskip}
   \hline
   \noalign{\smallskip}
   \multicolumn{5}{l}{Spectroscopy} \\
   \noalign{\smallskip}
   \hline
   \noalign{\smallskip}
   IRSPEC (3)   & 8.5,10.0,12.0	&  26	         &  264.8$\pm$ 9.6  & 0.054$\pm$0.004 \\
   PHT-S (12)   & 6.0...11.5 (0.5\,$\mu$m steps) &  89	& 296.5$\pm$ 8.3 & 0.043$\pm$0.002 \\
   CAM-CVF (15) & 7.5...14.5 (0.5\,$\mu$m steps) & 144	& 273.8$\pm$ 7.5 & 0.051$\pm$0.003 \\
   \hline
   \hline
   \noalign{\smallskip}
   \multicolumn{3}{c}{\bf Weighted Mean values:} & 273.0$\pm$11.9 & 0.050$\pm$0.005 \\
   \noalign{\smallskip}
   \hline
  \end{tabular}
  \end{center}
  \end{table*}
  
  Based on TPM parameters from Sect.~\ref{sec:tpm_input}, we derived
  radiometric effective diameter and albedo values from all
  photometric and spectroscopic measurements in
  Sect.~\ref{sec:obs}. The results are given in Table~\ref{tbl:tpm_res}.
  
  The typical observational and calibrational errors of individual measurements
  are between 5 and about 20\,\% in flux density. The translation into diameter leads
  to 3 to 10\,\% uncertainties for individual radiometric diameters.
  A 10\,\% change in radiometric diameter for \object{65~Cybele} requires an
  albedo modification of 0.010 (a larger diameter requires a lower albedo and
  vice versa). The $D_{eff}$ values per instrument are within $\pm$8\,\%
  of the weighted mean, only the UKIRT value is significantly off for
  unknown reasons. The ISOPHOT data were taken under the same aspect angles as
  the UKIRT data, but do not show such high $D_{eff}$ values.
  The highest quality photometric data are the CAM filter measurements.
  The derived diameter and albedo values are very close to the weighted mean
  values of all 10 sets of independent observations.

  The effective diameters are connected to the shape model
  via the equal volume sphere formula $D_{eff} = 2(abc)^{1/3}$
  (see also M\"uller \& Lagerros \cite{mueller98}). Shape, object
  orientation (represented by spin vector and aspect angle\footnote{Angle between
  the rotation axis of \object{65~Cybele} and the radius vector to Earth}), illumination,
  rotation and thermal effects are taken into account in the calculation
  of $D_{eff}$ and $p_V$.
  The calculated weighted mean effective diameter of 273.0\,km corresponds to an absolute
  ellipsoidal size of $2a \times 2b \times 2c = $ 301.9$\times$290.3$\times$232.2\,km
  for the given spin vector solution (first line in Table~\ref{tbl:sv}).

\subsection{Spectroscopic Analysis} \label{sec:spectroscopic}

  The original idea of the spectroscopic observations was to get
  new insides into the chemical composition and mineralogy of this main-belt
  asteroid. Mainly the nature of silicate features can be studied
  in the wavelength range between 8 and 12\,$\mu$m (Dotto et
  al.\ \cite{dotto00}).
  Green et al.\ (\cite{green85}) found no prominent
  emission features in the N-band spectrum of \object{65~Cybele}.
  
  \begin{figure}[btp!]
  \rotatebox{90}{\resizebox{!}{\hsize}{\includegraphics{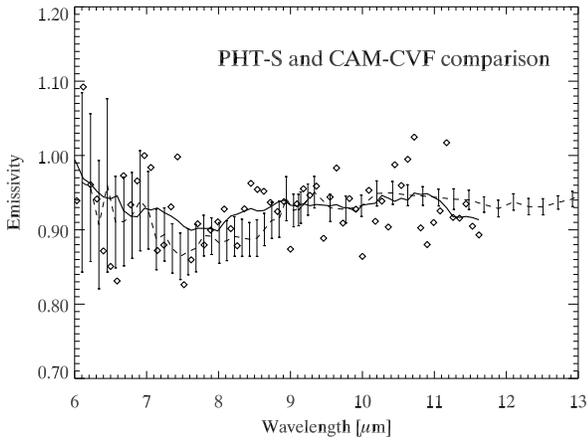}}}
  \caption{Comparison between PHT-S (diamonds) and CAM-CVF (dashed line
           with error bars) spectra divided by the corresponding TPM
           predictions with $\epsilon=1.0$. The solid line was produced
           by a 10-element smoothing of the PHT-S emissivity values.
           The CAM-CVF is the average of the up and down scans divided
	   by the model prediction.
     \label{fig:cam_phts_emi1}}
  \end{figure}

  Dotto et al. (\cite{dotto02}) and Barucci et al. (\cite{barucci02})
  analysed the spectroscopic features by dividing the observed spectrum by the
  corresponding TPM prediction under the assumption of a perfect
  emissivity at all wavelength ($\epsilon=1.0$). Here, we follow this
  procedure. Figure~\ref{fig:cam_phts_emi1} shows the PHT-S and the
  CAM-CVF measured spectrum divided by the corresponding TPM with
  $\epsilon=1.0$. Both spectra agree in spectral shape within the
  given observational and instrumental uncertainties.

  The PHT-S spectrum
  (Fig.~\ref{fig:cyb_phts_tpm}) was derived from 64 individual pixels
  which have a completely independent calibration for each pixel.
  The PHT-S data are therefore not ideal for an analysis
  of very low level spectroscopic features. 
  The CAM-CVF scans are much better suited for such an analysis.
  The source is always seen by the same pixels
  and with the variable filters the source spectrum is scanned. 
  Despite the disadvantages of the PHT-S spectrum for the analysis
  of shallow features, a 10 element smoothed PHT-S spectrum resembles
  the CVF spectrum very well.

\section{Discussion} \label{sec:discussion}

  \begin{figure}[btp!]
  \rotatebox{90}{\resizebox{!}{\hsize}{\includegraphics{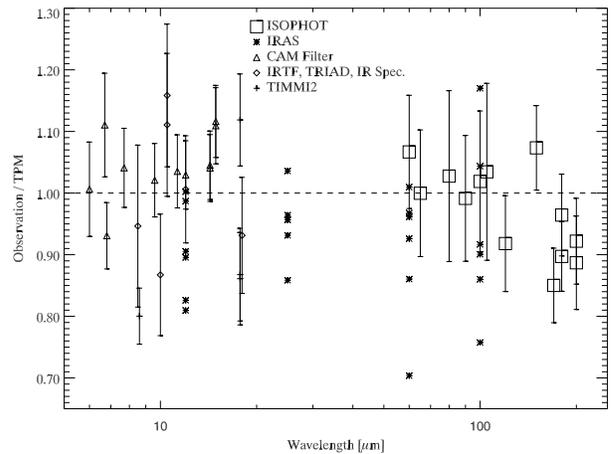}}}
  \caption{The observation over TPM ratios for the photometric
           measurements plotted against the corresponding wavelengths
           (see Fig.~\ref{fig:cyb_wavelength_stm} for STM ratios).
           The IRAS values are given without error bars for clarity.
           All TPM calculations were done with the resulting weighted
           mean values from Table~\ref{tbl:tpm_res}.
     \label{fig:cyb_wavelength}}
  \end{figure}

  \begin{figure}[btp!]
  \rotatebox{90}{\resizebox{!}{\hsize}{\includegraphics{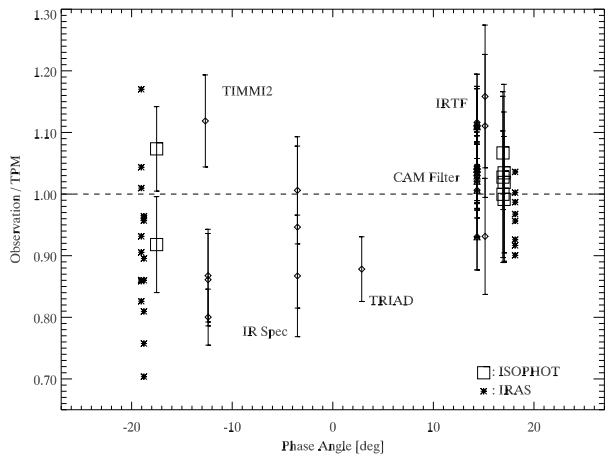}}}
  \caption{The observation over TPM ratios for the photometric
           measurements plotted against the corresponding phase angles.
           All TPM calculations were done with the resulting weighted
           mean values from Table~\ref{tbl:tpm_res}.
     \label{fig:cyb_phase}}
  \end{figure}

\subsection{Physical parameters}
\label{sec:dis_physical}

  Our reanalysis of 10 independent data sets (see Sect.~\ref{sec:obs})
  give now very consistent radiometric result through the TPM.
  We determined the effective diameter to D$_{eff}$=273.0$\pm$12.0 
  and the albedo to p$_V$=0.050$\pm$0.005. The discrepancy to the
  Tedesco et al.\ (\cite{tedesco02}) results is mainly
  related to the spherical shape simplification of the STM. Thermal
  effects, the beaming model, phase angle corrections and band-to-band
  corrections also contribute to the deviations. The occultation diameter
  was derived from a circle fitted to 3 measured chords. The uncertainties
  of the measurements and the limited information from a 3-chord 
  occultation event also allow for a fit with a larger slightly elongated
  ellipse, corresponding to the projected cross section of our derived
  radiometric size values. The fact that different data sets, taken
  at different wavelengths, different observing and illumination
  geometries give now one consistent solution is certainly in favor
  of our larger diameter. And, all IRAS observations fit nicely into that
  picture, including the 100\,$\mu$m data which were not used for
  the Tedesco et al.\ (\cite{tedesco02}) calculations. Here, we would
  like to emphasize that we did not apply any band-to-band corrections
  or any other wavelengths-, phase angle or beaming corrections,
  like it was done for the STM IRAS applications.

  Despite the usage of best available ellipsoidal shape model and the
  nice agreement on average between the TPM and the observations,
  we believe that the unknowns of the true shape limit the accuracy
  of TPM predictions for individual epochs. Deviations between the
  true object cross section and the projected model cross section are
  easily possible, but so far no improved shape model
  (e.g.\ from radar or direct imaging methods) is available.

  The shape model is only useful together with the spin vector solution.
  Erikson (\cite{erikson00}) gives 2 solutions as result of lightcurve
  inversion methods (see Table~\ref{tbl:sv}), but both solutions have
  the same sense of rotation. The determination of the correct solution
  on basis of radiometric analysis is therefore very difficult (see
  Sect.~\ref{sec:tpm_input} for details).
  Here, we accepted solution 1 for all calculations.

\subsection{Thermal parameters}
\label{sec:dis_thermal}

  We used a flat emissivity model with $\epsilon=0.9$ over all wavelengths
  out to 100\,$\mu$m, at longer wavelengths the model emissivity is slowly
  decreasing towards values of 0.8 at sub-mm wavelength.
  Previously published emissivity values of 1.0 at around 20\,$\mu$m
  were based on ground-based Q-band observations and IRAS 25\,$\mu$m
  fluxes together with very uncertain beaming values
  (M\"uller \& Lagerros \cite{mueller98}). With our modification 
  of taking a constant 0.9 value at mid-IR wavelengths we investigate
  the emissivity behaviour in this critical region where beaming
  effects play an important role (Dotto et al.\ \cite{dotto00}).
  In Fig.~\ref{fig:cyb_wavelength} there is no trend with wavelength
  visible out to about 150\,$\mu$m. The constant `observation
  over TPM' ratios in Figs.~\ref{fig:cyb_phts_tpm} and \ref{fig:cyb_cam_tpm}
  confirm our assumption of a constant $\epsilon=0.9$ at mid-IR wavelengths.
  The 5 ISOPHOT measurements at wavelength beyond 150\,$\mu$m
  (calibrated against a few stars and the planets \object{Uranus} and
  \object{Neptune}) show fluxes which are about 10\,\% lower than the TPM predictions
  (Fig.~\ref{fig:cyb_wavelength}). 
  M\"uller \& Lagerros (\cite{mueller02}) see nice agreement with this
  current emissivity model for \object{1~Ceres}, \object{2~Pallas},
  \object{4~Vesta} and \object{10~Hygiea} out to 200\,$\mu$m,
  although for \object{2~Pallas} and \object{4~Vesta} there were indications
  for a similar trend
  at the very longest wavelengths. The observed far-IR fluxes are
  to a good approximation directly proportional to the hemispherical
  model emissivity (M\"uller \& Lagerros 1998). Instead of a slow decrease
  from 0.9 at 100\,$\mu$m to 0.8 at 400\,$\mu$m, the \object{65~Cybele} observations
  now indicate a more abrupt decrease in emissivity to values around 0.7-0.8
  in the wavelength range 150 to 200\,$\mu$m.

  Redman et al. (\cite{redman92}; \cite{redman98}) found effective emissivities
  between 0.5 and 0.9 in the sub-mm range for 7 large asteroids. They
  attribute differences in the emissivity behaviour to different optical
  depth of the warm surface material and its density. Scattering processes
  by grains within the regolith reduce the emissivity in a wavelength
  dependent fashion. For the Moon the far-IR emission is also suppressed
  by scattering processes (Simpson et al. \cite{simpson81}). Scattering
  processes are most effective if wavelength and grain sizes have similar
  dimensions.  Following this interpretation, our far-IR low emissivities indicate
  that the dominant particle size in the \object{65~Cybele}'s regolith (and maybe also for
  \object{2~Pallas} and \object{4~Vesta}) would be around 200\,$\mu$m.
  More observations are needed to identify the precise emissivity behaviour.
  Well-calibrated spectroscopic observations in the far-IR and sub-millimetre
  (e.g.\ with HERSCHEL) will then allow a detailed description of particle
  size distributions within the regolith.

  The measurements before and after opposition come from different
  instruments and are too inhomogeneous to identify clear trends with
  phase angle, especially as the measurements before opposition 
  (negative phase angle) have too large error bars.
  We varied the thermal inertia between 0 and
  50\,$\mathrm{J\,m^{-2}\,s^{-0.5}\,K^{-1}}$ to study the systematic changes
  in the data representation of Fig.~\ref{fig:cyb_phase}. The zero thermal
  inertia produces a significantly larger scatter in the observation-to-model
  ratios. The $\Gamma=50$ calculations showed
  a very pronounced asymmetry between before and after opposition
  ratios. Our default value of 15\,$\mathrm{J\,m^{-2}\,s^{-0.5}\,K^{-1}}$
  seems to work fine, but somewhat smaller or larger values cannot
  be excluded. The much larger scatter in the before opposition ratios
  is pure coincidence. E.g.\ the before opposition IRAS data cover
  a much larger range in flux and the most deviating IRAS ratios
  on the left side of Fig.~\ref{fig:cyb_phase} belong to the lowest
  measured fluxes.

  The beaming model with $\rho=0.7$ and $f=0.6$ influences strongly the
  slope of the observation to model ratios in the mid-IR (Dotto et al.\
  \cite{dotto00}). The flat curves in Figs.~\ref{fig:cyb_phts_tpm} and 
  \ref{fig:cyb_cam_tpm} confirm our ``default beaming model". About 10-20\,\%
  modifications of the r.m.s.\ slope value ($\rho$) and the fraction of surface
  covered by craters (f) would still be in agreement with the PHT-S and
  the CAM filter measurements.
  M\"uller (\cite{mueller02a}) demonstrated in different
  model simulations the influences of the thermal parameters
  with wavelength and phase angles.
  But thermal inertia, beaming effects and the low level
  mineralogic emission features play together and influence
  the slope of the thermal emission in a complicated way
  and the different contributions are not easy to disentangle.

\subsection{Emissivity analysis}
\label{sec:dis_emissivity}

  \begin{figure}[btp!]
  \rotatebox{90}{\resizebox{!}{\hsize}{\includegraphics{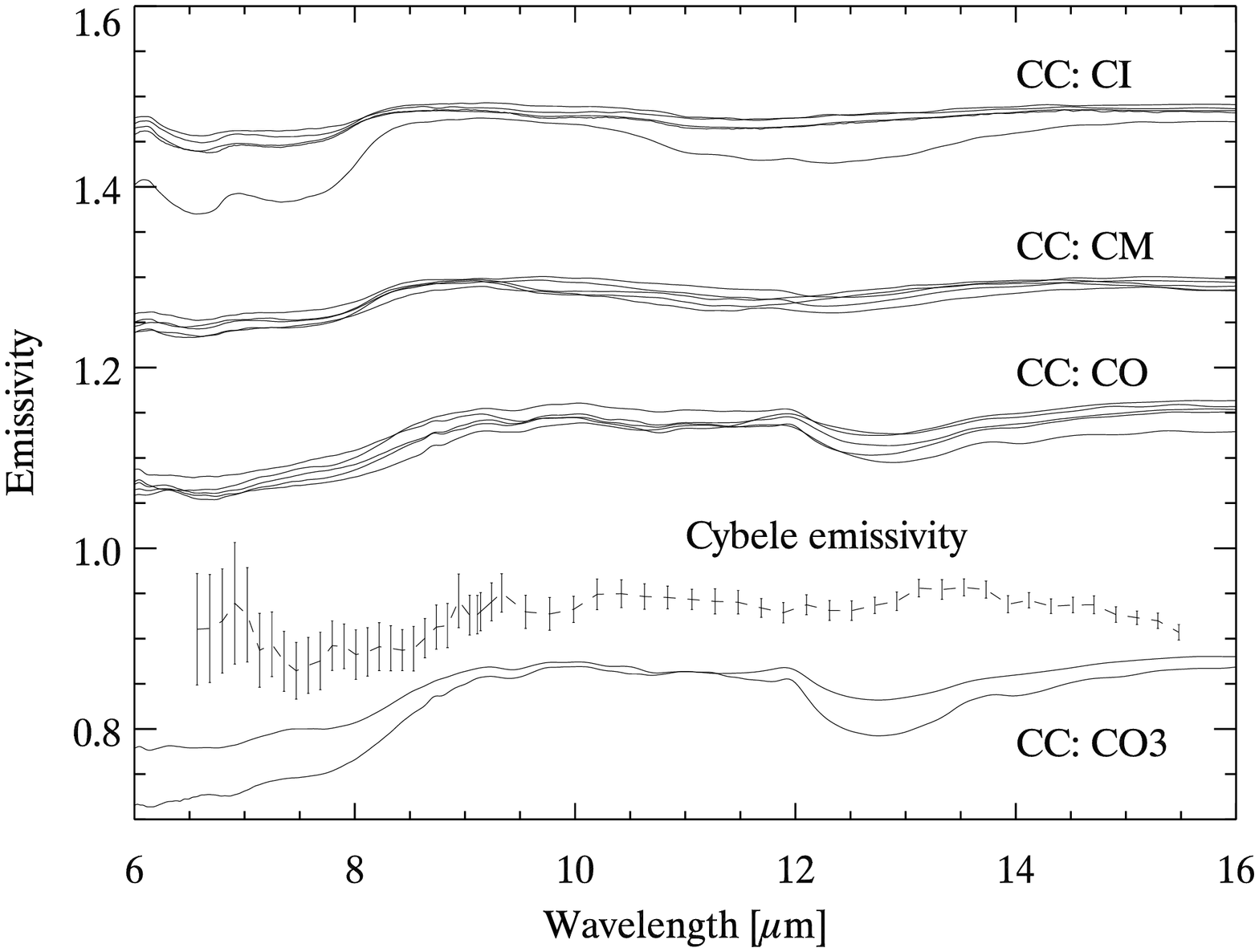}}}
  \caption{The average CAM-CVF scan was divided by a TPM prediction with model
           emissivity set to 1.0 at all wavelengths. The average CVF/TPM ratio
           is plotted as dashed line. The \object{65~Cybele} emissivity curve
           shows no prominent features in the 7 to 15\,$\mu$m range.
           Different carbonaceous chondrites are shown for comparison
           and vertically shifted for clarity.
     \label{fig:cvf_meteorites}}
  \end{figure}

  Barucci et al.\ (\cite{barucci87}) classified \object{65~Cybele} as
  C0-type asteroid on basis of visual multi-color and IRAS data.
  They associated this taxonomic type with carbonaceous
  chondrites. The asteroids \object{10~Hygiea} and \object{511~Davida},
  also C0-types, were found to show similarities in near-IR and 3\,$\mu$m
  spectroscopy with carbonaceous chondrite meteorites of CI/CM subclass
  (Hiroi et al.~\cite{hiroi96}).
  Therefore, we concentrated our search for meteoritic analogues to
  carbonaceous chondrites, which represent, chemically and mineralogically,
  the most primitive material in the solar system.
  
  In Fig.~\ref{fig:cvf_meteorites} we show the average CVF scan (divided
  by a TPM prediction with $\epsilon=1.0$) with the error bars derived
  from the two CVF scans ($\sigma_{mean}/\sqrt{2}$).
  For comparison, we plotted the emissivity of different
  powdered stony meteorite samples, representing 17 carbonaceous chondrite
  meteorites at small particle sizes of 0--75\,$\mu$m (Salibury et
  al. \cite{salisbury91a}; \cite{salisbury91b}, ASTER spectral
  library\footnote{http://speclib.jpl.nasa.gov}). This library contains
  60 meteorite samples all at grain sizes of 0-75\,$\mu$m.
  The sample includes five meteorites of subclass CI, five of subclass CM,
  five of subclass CO and two of subclass CO3.
  The emissivity spectra (1 -- measured reflectivity) are offset for clarity.

  The \object{65~Cybele} emissivity shows no prominent features, but
  an increase in emissivity between 8.0 and 9.5\,$\mu$m
  is seen in both down and upward CVF scans and is thus a reliable feature,
  contrary to the part below 7.0\,$\mu$m where the source becomes weak and 
  transient problems hamper the measurement (see Sect.~\ref{sec:cvf}).
  Between 12.5 and 14\,$\mu$m we recognise a very shallow feature which
  is also present in both scans.

  Salibury et al. (\cite{salisbury91a}) noted that a rather steep
  increase in emittance, the so-called Christiansen feature (CF), occurs
  between the wavelength region dominated by volume scattering
  (roughly below 7.5\,$\mu$m) and the wavelength region dominated
  by surface scattering and reststrahlen bands (roughly beyond 8.5\,$\mu$m).
  They placed the CF wavelengths of different types of carbonaceous
  chondrites within the region between about 8.7 and 9.4\,$\mu$m.
  The CO and CO3 types of the carbonaceous (Fig.~\ref{fig:cvf_meteorites})
  come closest in CF similarity with the \object{65~Cybele} emittance peak between
  9.0 and 9.4\,$\mu$m, but the data quality is not sufficient for a
  conclusive identification.
  The two CO3 stony meteorites in Fig.~\ref{fig:cvf_meteorites} are
  Isna (upper curve) and ALHA~77003 (lower curve), the CO meteorites
  are Lance (2$\times$), ALH~83108, Ornans and Warrenton (Salibury et
  al. \cite{salisbury91a}). The CI and CM subclasses show the CF at
  slightly shorter wavelengths.

  Our results from the emissivity analysis (Sect.~\ref{sec:dis_thermal})
  pointed at predominantly larger particles as cause for the
  far-IR emissivity drop. But the Salibury et al. (\cite{salisbury91a})
  library provides only spectra of small grain sizes.
  Le Bras \& Erard \cite{lebras03} found that the grain size effect
  can shift the CF in wavelengths and change also the contrast.
  Meteoritic samples with larger grain sizes are therefore needed
  to proove direct connections between the measured spectroscopic features
  on asteroids and meteoritic samples.

  For the feature between 12.5 and 14\,$\mu$m we found no counterpart yet,
  but phyllosilicates, like Kaolinite, Lepidolite or Serpentine, show
  emissivity increases at around 12.5\,$\mu$m, similar to
  what we see here. The fine-grained phyllosilicates are one main
  component of CI/CM carbonaceous chondrites, while the CO types are
  anhydrous materials. This would be an argument in favour of similarities
  with CI or CM types, but again, laboratory studies at larger grain
  sizes are needed to clarify the situation.

  Figure~\ref{fig:cvf_meteorites} clearly reveals the difficulties
  of identifying mineralogic features based on mid-IR spectroscopy.
  Laboratory spectra of different meteorites at various 
  grain sizes would be necessary.
  Concerning the observations of asteroids, a high S/N ratio over
  a large wavelengths range is necessary to detect the low level
  emissivity changes.
  These features would be very difficult to detect from
  ground-based N-band spectroscopy (8 to 13\,$\mu$m) where
  atmospheric effects limit the wavelength range and degrade the data quality.
  Additional complications arise from the strong slope changes
  around the thermal continuum energy peak at around 14\,$\mu$m.
  Poor spectral information together with wrongly
  modelled continuum slopes can then easily lead to misinterpretation
  at the band borders.
  The modelling of the surface temperature distribution, the thermal
  inertia as well as the beaming parameters $\rho$ and
  $f$ influence the slope of the thermal emission. Our thermal values
  (Sec.~\ref{sec:tpm_input}) worked perfectly for the PHT-S and
  the CAM-CVF data sets, but the spectroscopic interpretation still
  remains difficult.
  
\subsection{Comparison with the STM}
\label{sec:dis_stm}

\begin{figure}[btp!]
  \rotatebox{90}{\resizebox{!}{\hsize}{\includegraphics{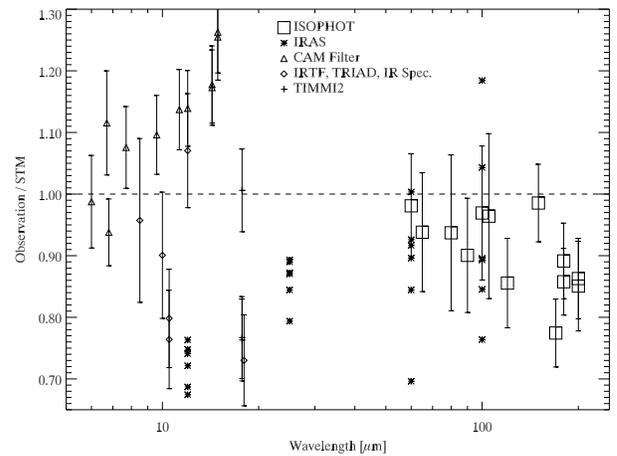}}}
  \caption{The observation over STM ratios for the photometric
           measurements plotted against the corresponding wavelengths
           (see Fig.~\ref{fig:cyb_wavelength} for TPM ratios).
           All STM calculations were done with the resulting weighted
           mean values from Table~\ref{tbl:tpm_res}. 
     \label{fig:cyb_wavelength_stm}}
  \end{figure}

  The shape and spin vector information is very important for
  a consistent analysis of the combined observations. Although
  we only used a simple ellipsoidal shape model, one
  can see the shape influence by comparing Fig.~\ref{fig:cyb_wavelength_stm}
  with Fig.~\ref{fig:cyb_wavelength} in Sect.~\ref{sec:dis_physical}.
  For Fig.~\ref{fig:cyb_wavelength_stm} we calculated the model values
  via the STM (Lebofsky et al.\ \cite{lebofsky86}), using a spherical shape model.
  The scatter between data sets taken under different
  aspect angles is much larger. Some measurements which agree in
  Fig.~\ref{fig:cyb_wavelength} are now offset by large amounts
  (e.g.\ the IRAS 12 and 25\,$\mu$m observations and the CAM filter
  measurements). Assuming a spherical shape (with the given spin vector
  and thermal behaviour) leads to a weighted mean diameter of 280.0\,km
  and an albedo of 0.047. But now the standard deviations would increase
  by about 60\,\% for the diameter and about 50\,\% for the albedo!
  This result clearly demonstrates the importance of using spin vector
  and shape information for combining observations from different epochs.
  Modifying the STM beaming value might improve the ``Observation/STM"
  ratios for some observations, but different observations would require
  then different STM beaming values. And, for some data sets the STM would
  require a wavelength dependent $\eta$ to account for the trends in
  Fig.~\ref{fig:cyb_wavelength_stm}.

  \begin{figure}[btp!]
  \rotatebox{90}{\resizebox{!}{\hsize}{\includegraphics{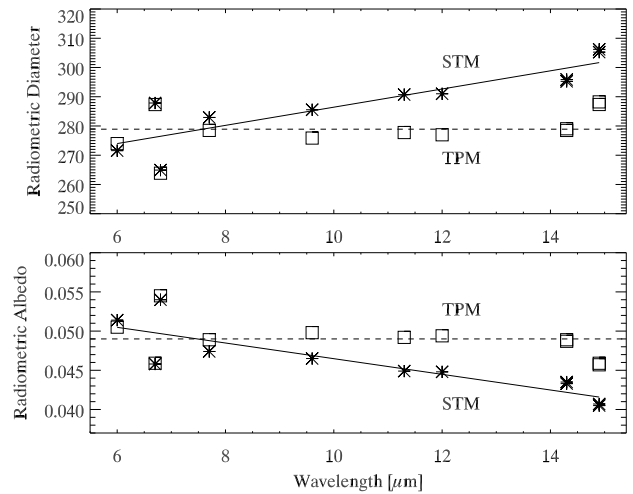}}}
  \caption{The derived diameters and albedos for the CAM filter
           measurements as a function of wavelength, for the TPM
	   and also the STM.
     \label{fig:tpm_stm}}
  \end{figure}

  We used the most reliable data set --the CAM filter measurements-- 
  for a comparison between TPM and STM radiometric results.
  The TPM analysis leads to a constant diameter and albedo value for all
  CAM filter measurements, while the STM produces wavelength dependent
  values (Fig.~\ref{fig:tpm_stm}). The albedo values show the
  opposite behaviour. The derived
  diameters show no trend with wavelengths (see also Fig.~\ref{fig:cyb_cam_tpm},
  lower part for a comparison of the observation to TPM ratios over wavelengths)
  and agree within 2-3\,\% (about 6\,\% for the albedo).
  This also confirms our simplified constant emissivity values in this
  wavelength range close to the thermal emission peak. Simulations of the
  \object{65~Cybele} constellation during the CAM filter epoch show that this wavelength
  dependent trend continues out to about 30\,$\mu$m. At longer wavelengths
  also the STM would produce an almost constant diameter. At 30\,$\mu$m the
  STM diameter would be about 10\,\% larger. This wavelength dependent
  effect of the STM is mainly produced by the STM correction factor
  for infrared beaming. Detailed studies of the beaming influence with
  wavelengths and phase angles can be found in M\"uller (\cite{mueller02a}).

  Figures~\ref{fig:cyb_wavelength}, \ref{fig:cyb_phase} and 
  \ref{fig:cyb_wavelength_stm} reveal additional problems of the STM.
  The scatter in the observed fluxes divided by model calculations is much larger
  for the STM (Fig.~\ref{fig:cyb_wavelength_stm}).
  The main reason for the larger scatter is the assumption
  of a spherical shape. But also the simplified beaming, emissivity and
  phase angle corrections can be recognized in Figs.~\ref{fig:cyb_wavelength}
  and \ref{fig:cyb_wavelength_stm}. The residual beaming effects cause wavelengths
  dependent ratios in Fig.~\ref{fig:cyb_wavelength_stm},
  easily visible for the IRAS data ($\star$-symbol) or the CAM filter
  data ($\triangle$-symbol). The offset between these 2 sets is caused
  by the wrong cross section due to the spherical shape model of STM.
  The STM calculation of the ratios of Fig.~\ref{fig:cyb_phase} 
  shows a strong asymmetry between the ratios before and after
  opposition. Additionally, the scatter between the different data
  sets would be much larger. Unfortunately, the overlaying shape effects 
  of the inhomogeneous data did not allow to compare the 0.01\,mag/$^{\circ}$
  phase angle correction factor of the STM with the more sophisticated
  calculations of the true illumination and observing geometry by the TPM.

\section{Conclusions} \label{sec:conclusion}

  The recently developed TPM allows 
  to combine thermal infrared observations with information from
  visual photometry, lightcurves, close-up observations or direct
  measurements. Using \object{65~Cybele} as a ``text book example", we applied
  this powerful TPM in several different ways. First, we used
  re-calibrated observations from literature and recently obtained
  thermal infrared measurements to derive the size and albedo of \object{65~Cybele},
  taking also shape and spin vector information into account.
  Our derived effective diameter (273.0$\pm$11.9\,km) and
  albedo values (0.050$\pm$0.005) deviate significantly from
  the IRAS based results of 237.3$\pm$4.2\,km and
  0.071$\pm$0.003 (Tedesco et al.\ \cite{tedesco02}). We attribute
  these differences to the limitations of the STM 
  (Sects.~\ref{sec:dis_physical}, \ref{sec:dis_thermal} and \ref{sec:dis_stm}).

  The spectroscopic analysis confirmed the proposed ``default
  thermal behaviour" (M\"uller et al.\ \cite{mueller99}) for
  large main-belt asteroids which is dominated by a surface
  regolith with very low thermal inertia. The related beaming model
  and thermal inertia values are the key elements to match
  even the high accuracy ISO measurements at the critical
  mid-IR range. In return, reliable diameter and albedo
  values can be determined from individual filter measurements
  at any thermal infrared wavelength. No band-to-band or
  wavelength dependent correction factors are necessary anymore.
  Additionally, the thermal emission modelling provides the
  basis for the analysis of mineralogic features and allows
  to investigate relations to meteoritic samples.

%
%
%


\begin{acknowledgements}
We would like to thank U.\ Klaas and P.\ Abraham for their support in the
PHT-S data analysis and D.\ Osip for the provision of the IRTF observations.
R.\ Siebenmorgen and M.\ Sterzik supported the ESO observing run
and the data analysis of the TIMMI2 measurements.
The ISOPHOT data presented in this paper were reduced using PIA, which is
a joint development by the ESA Astrophysics Division and the ISOPHOT
Consortium with the collaboration of the Infrared Processing and Analysis
Center (IPAC). Contributing ISOPHOT Consortium institutes are
DIAS, RAL, AIP, MPIK, and MPIA.
The ISOCAM data presented in this paper were analysed using `CIA', a
joint development by the ESA Astrophysics Division and the ISOCAM
Consortium. The ISOCAM Consortium is led by the ISOCAM PI, C.\
Cesarsky.
We also thank the referee Dr.\ Delbo for his careful review.
\end{acknowledgements}


\end{document}